\documentclass[aps,nofootinbib,twocolumn,prl,pacs]{revtex4} 
\usepackage{amsfonts}
\usepackage{amssymb}
\usepackage{graphicx}
\usepackage{pstricks}
\usepackage{epsfig}
\newcommand{\be}{\begin{equation}}
\newcommand{\ee}{\end{equation}}
\newcommand{\bear}{\begin{eqnarray}}
\newcommand{\eear}{\end{eqnarray}}

\newcommand{\lapproxeq}{\lower .7ex\hbox{$\;\stackrel{\textstyle
<}{\sim}\;$}}
\newcommand{\gapproxeq}{\lower .7ex\hbox{$\;\stackrel{\textstyle
>}{\sim}\;$}}
\newcommand{\stackdown}[2]{\lower 1.4ex\hbox{$\;\stackrel{\textstyle{#1}}
{\scriptstyle{#2}}\;$}}
\newcommand{\beq}{\begin{equation}}
\newcommand{\eeq}{\end{equation}}

\makeatletter
\def\slash{\@ifnextchar[{\fmsl@sh}{\fmsl@sh[0mu]}}
\def\fmsl@sh[#1]#2{%
  \mathchoice
    {\@fmsl@sh\displaystyle{#1}{#2}}%
    {\@fmsl@sh\textstyle{#1}{#2}}%
    {\@fmsl@sh\scriptstyle{#1}{#2}}%
    {\@fmsl@sh\scriptscriptstyle{#1}{#2}}}
\def\@fmsl@sh#1#2#3{\m@th\ooalign{$\hfil#1\mkern#2/\hfil$\crcr$#1#3$}}
\makeatother

\begin{document}
\title{Relativistic Modified Newtonian Dynamics from String Theory?}
\author{Nick E.~Mavromatos and Mairi Sakellariadou}
\affiliation{King's College London,
Department of Physics, Strand WC2R 2LS, London, U.K.}

\begin{abstract}
We argue that TeVeS-like vector fields appear naturally in certain
string theory backgrounds involving D0-branes, as a result of the
recoil velocity field, expressing the interaction of neutrino string
matter with point-like branes. However, the similarity with TeVeS
models is restricted only to the bi-metric properties of space time,
namely the difference of the background metric from the one felt by
(some) matter fields interacting, in a topologically non-trivial
manner, with the D0-brane defects.  In our approach, neutrinos appear
as dark matter candidates that could be ``captured'' by the D0 branes,
as a result of stringy properties, and thus couple with the
recoil-vector fields.  Moreover, we argue in support of a possibly
preferential r\^ole of neutrinos in inducing novel non-perturbative
contributions to ``vacuum'' (dark) energy, in addition to their
ordinary dark matter contribution. In fact, the r\^ole of neutrinos as
providing substantial contributions to dark matter and dark energy
components of the Universe, suggested by our approach, appears also to
be necessitated by the need to reproduce the peaks in the Cosmic
Microwave Background radiation spectrum, as claimed recently in the
literature. Thus, our framework may be viewed as providing a
microscopic explanation of such phenomenological conclusions
concerning TeVeS-like, Lorentz-violating models.

\vspace{.2cm}
\end{abstract}

\maketitle


Overwhelming experimental evidence from diverse sources~\cite{observ},
appear to provide convincing arguments in support of the existence of
$70 \%$ of the universe's total density in the form of dark energy,
the fuel that drives acceleration, and $23\%$ in an unknown form of
weakly interacting Dark Matter (DM).  However, the above studies are
not model independent, as they are based on conventional
Einstein/Friedmann-Lema\^{i}tre-Robertson-Walker (FLRW) cosmologies.
Deviations from these assumptions may lead to a different picture
concerning the various contributions to the current-epoch energy
budget of the universe.  For instance, in non-equilibrium string
theory models~\cite{diamandis}, it is possible to fit the data 
 with the conventional $\Lambda$CDM model~\cite{emmncosmo}.  The
important feature of these models is that the contributions to the
universe's energy budget, are coming from both dark energy and (dark
and ordinary) matter terms.

More drastic conclusions can be drawn on the nature of the
``dark sector'' of the universe if one is prepared to abandon the
Einstein gravity theory at galactic scales, in favour of a MOdified
Newtonian Dynamics (MOND)~\cite{MOND} below some universal
gravitational acceleration scale $a_0$. In such a case, the entire
concept of DM may not be needed.  Indeed, MOND theory, without DM at
galactic scales, appears capable of explaining the rotational curves
of galaxies, whose observed departure from the expected Keplerian law,
in case all matter were luminous, led initially to the DM postulate.
Although MOND originally appeared as a purely empirical (and not
entirely covariant) theory, nevertheless recently
Bekenstein~\cite{teves} proposed a relativistic field theory version
of it, involving TEnsor, VEctor and Scalar fields (TeVeS) of
gravitational origin, reproducing the dynamics of MOND.

The TeVeS theory contains two metrics: $g^E_{\mu\nu}$,
 the ``Einstein-frame'' metric, which
satisfies the canonically normalised Einstein-Hilbert action, and
 $g_{\mu\nu}$,  that matter ``feels''; all
geodesics are calculated in terms of $g_{\mu\nu}$. It is related
to the Einstein-frame metric by~\cite{teves}: 
\be g_{\mu\nu} =
e^{-2\Phi}g^E_{\mu\nu} + \left(e^{-2\Phi} - e^{2\Phi}\right)A_\mu
A_\nu~;
\label{mattermetric}
\ee $A_\mu $ ($\Phi$) is the vector (scalar) field of  the TeVeS
 theory.
The presence of a vector field, $A_\mu$, is crucial for consistency of
the TeVeS models. Isotropy and homogeneity of the universe are not
disturbed if $A_\mu$ has only a time-like component, which
depends only on the cosmic time of the co-moving frame $\tau$, namely
$A_\mu = \left(A_0 (\tau ), 0, 0, 0 \right)$,
while internal consistency of the TeVeS theories implies a constraint
on the fields magnitude~\cite{teves}: \be A_\mu A^\mu = - 1~,
\label{constrvec}
\ee where the indices are contracted with respect the Einstein-frame
metric.  This constraint, of course, leads to Lorentz violation in the
sense of its direction defining a ``preferred frame'', thereby
constituting a modern version of (isotropic)
``aether''~\cite{aether}. In a cosmological framework, with a scale
factor $a(\tau)$, and Einstein metric such that $g_{00}^E =
-a^{-2}(\tau)e^{2\Phi (\tau)}$, the constraint Eq.~(\ref{constrvec})
implies that 
$ A_0 (\tau) = a(\tau) e^{\-\Phi (\tau)}.$ 
 The action of $A_\mu$
is described by a Maxwell-like
kinetic term, in terms of its field strength ($F_{\mu\nu}=\partial_\mu
A_\nu - \partial_\nu A_\mu$) plus a Lagrange multiplier term
implementing the constraint, Eq.~(\ref{constrvec}). The coefficient of the
Maxwell term in the action is a  free parameter.

In TeVeS the dynamics of  $\Phi$ is described by an
action including a non minimal coupling of the kinetic terms of $\Phi$
with another, non dynamical, scalar field $\mu$, with a potential
$V(\mu)$ whose form determines the phenomenology as far as the
reproduction of the observed rotational curves of galaxies is
concerned~\cite{teves}.  In Ref.~\cite{skordis} it was argued that
TeVeS can reproduce the observed galactic power spectrum of baryonic
fluctuations. This was a prediction of the DM hypothesis: just a
baryonic universe predicts pronounced wiggles in the power spectrum
$\Delta \equiv k^3 P(k)/2\pi^2$, which are not observed.  However,
Ref.~\cite{skordis} argued that TeVeS could also explain the absence
of the wiggles and reproduce the observed galaxy spectra without 
 DM.  It is the vector field of TeVeS that plays a
crucial r\^ole in reproducing the observed spectra~\cite{liguori}.

The arguments on the r\^ole of the vector field in reproducing large
scale structures and the correct phenomenology seem to be generic, and
not depending on the detailed Lagrangian of the original TeVeS theory.
Only the basic important features of the theory, namely the existence
of the two metrics and of the (Lorentz-violating) ``aether-like''
vector field, appear to be important in this respect.  TeVeS can also
reproduce~\cite{skordis} the observed positions of acoustic peaks in
the CMB spectrum, only if a contribution from \emph{massive neutrinos}
is included, with a current density $\Omega_\nu \sim 0.15-0.17 $, and
masses $\sim 2$ eV, together with a \emph{cosmological constant}
contribution $\Omega_\Lambda \sim 0.80 - 0.78$, respectively, and
ordinary (primarily baryonic) matter $\Omega_{\rm b}\sim 5\%$.
However, TeVeS theories have some drawbacks.  They appear as
``phenomenological'', not derived from an underlying microscopic
theory, and they fail to explain the behaviour observed in certain
systems, such as the bullet cluster~\cite{bullet}, or galaxies
claimed to be dominated by ``dark matter''~\cite{onlydm}.

All the above features call for a microscopic explanation.  Our aim
is to argue in favour of a link between them, in the framework of some
specific backgrounds of string theory~\cite{emn}.  The model involves
two stacks of parallel eight-dimensional brane worlds embedded in a
nine-dimensional bulk space. The bulk space-time is restricted in a
specific way by two orientifolds, whilst the brane worlds can be
compactified to three (spatial) dimensions in a specific way, whose
details are not important here.  The bulk involves D0-branes
(D-particles), which can propagate in \emph{both} the bulk and the
brane worlds.  They are viewed as point-like \emph{defects} on the
space-time.  The D-particles have mass $m_{\rm D} = M_{\rm s}/g_{\rm
s}$ ($M_{\rm s}$ the string scale and $g_{\rm s} < 1$  the
string coupling).  If $M_{\rm s}$ is at a TeV scale, then the mass of the
defect is heavier. There are models in which $M_{\rm s} =M_{\rm Pl}$
(the four-dimensional Planck scale), in which case the D-particles
have trans-Planckian mass.

One of the (compactified) three-branes plays the r\^ole of our
observable universe, on which Standard Model matter lives. The latter
is described by open strings ending to the D-brane world. Closed
strings describe excitations in the gravitational multiplet of
strings, and propagate in \emph{both} the bulk and the brane.  When
the branes and the D-particles are \emph{at rest} the vacuum is
supersymmetric, with \emph{zero vacuum energy}. Motion of branes
breaks the target-space supersymmetry resulting in non -trivial
contribution to the vacuum (``dark'') energy.  Interaction of open
strings on the brane with D-particles can be described by logarithmic
field theory of `recoil'/impulse deformations of the pertinent
$\sigma$-model describing stringy excitations on the D-particle
defect~\cite{recoil}. The interaction is not a smooth scattering
event. It involves the \emph{capture} of a string by the D-particle,
which formally imply the change of world-sheet boundary conditions
from Neumann to Dirichlet.  Electric charge conservation implies that
charged string excitations cannot be captured by the D-particle, due
to the electromagnetic gauge symmetry $U_{\rm em}(1)$.  Since the
D-particles also carry another $U(1)$ symmetry, which is unrelated to
electromagnetism, flux conservation for that symmetry as well implies
that an isolated D-particle, with a string emanating from it with one
free end in the bulk space, \emph{cannot} exist.  Thus, from the
elementary particles of the Standard Model, only electrically neutral
particles can interact with D-particles.

As we shall discuss later on, \emph{neutrinos} appear to have a
preferential r\^ole in inducing novel non-perturbative contributions
to the vacuum energy~\cite{vitiello,barenboim}, as compared to the
other electrically neutral excitations that could in principle
interact with D-particles.  During each capture process, the massive
D-particle defect recoils, and thus disturbs the surrounding
space-time, resulting in metric distortions $\delta g_{\mu\nu}$. Such
distortions are found by following conformal field theory methods on
the world-sheet of the pertinent $\sigma$-model~\cite{recoil}. The
vertex operator on the world-sheet boundary $\partial \Sigma$ reads
$$\int_{\partial \Sigma} u_i X^0 \Theta_\epsilon (X^0) \partial_n X^i~,$$
where $\partial_n$ denotes normal world-sheet derivative, $X^0$
($X^i$) obeys Neumann (Dirichlet) boundary conditions on the
world-sheet, $u_i = \gamma v_i = \gamma g_s \Delta k_i /M_{\rm s} $ is
the ``recoil'' three-velocity of the D-particle defect during the
capture process, $\gamma = (1 - {\vec v }^2 )^{-1/2}$ is the Lorentz
factor, and $\Delta k_i$ is the momentum transfer of the
stringy-matter excitation. The operator $\Theta_\epsilon (X^0)$,
$\epsilon \to 0^+$, is a regularised Heaviside function, denoting the
moment of impact, $X^0=0$ for definiteness~\cite{recoil}.  World-sheet
conformal invariance requires that a target-space metric deformation
corresponds to a conformal-dimension-two operator of the $\sigma
$-model. To find this deformation we first rewrite~\cite{recoil} the
boundary deformation as a total world-sheet derivative in a
target-space covariant form 
$$\int_\Sigma \partial_{\hat \alpha} \left(
u_\mu u_\nu X^\nu \Theta_\epsilon (u_\rho X^\rho) \partial^{\hat
\alpha} X^\mu \right)~,$$ where ${\hat \alpha} =1,2$ is a world-sheet
index, $u_\mu$ is a four-velocity, \be u_\mu u^\mu = -1~.
\label{fourvelconstr}
\ee
For a linear dilaton $\Phi = -u_\mu X^\mu$ and $\sigma$-model frame
Minkowski metric backgrounds~\cite{aben}, $e^{-\Phi}$ acts as the
``Liouville dressing operator''~\cite{ddk}, restoring conformal
invariance, {\it i.e.} 
$$\int_\Sigma e^{-\Phi} \partial_{\hat
\alpha} \left( u_\mu u_\nu X^\nu \Theta_\epsilon (u_\rho X^\rho)
\partial^{\hat \alpha} X^\mu \right)$$ has dimension two in the limit
$\epsilon \to 0^+$: \be V_{\rm bulk~rec} = \int_\Sigma e^{u_\mu X^\mu}
u_\mu u_\nu \partial X^\mu \partial X^\nu \Theta_\epsilon (X^0) +
\dots~,
\label{dressed}
\ee where the $\dots$ denote terms which vanish either upon using the
world-sheet equations of motion, or upon taking the limit $\epsilon
\to 0^+$.  Indeed, due to the existence of a linear dilaton
background, the operator $e^{-\Phi} = e^{u_\mu X^\mu}$ has conformal
dimension $u_\mu ( u^\mu - u^\mu ) = 0$, thus it does not affect the
overall conformal dimension two of the bulk recoil operator,
Eq.~(\ref{dressed}), as a result of the $\partial X^\mu \partial X^\nu
$ part.

The result of Eq.~(\ref{dressed}) points towards the existence of a
target-space deformation due to the D-particle recoil: 
$$ \delta
g_{\mu\nu} \propto e^{-\Phi} u_\mu u_\nu \Theta_{\epsilon \to 0}(u_\rho
X^\rho)~.$$ However, to ensure a smooth connection with the flat metric
at the origin of the boosted time $u_\rho X^\rho = 0$, it is necessary
to impose the condition that $\delta g_{\mu\nu}=0$ for $ u_\rho X^\rho
= 0$ (the reader should recall that at the co-moving frame of the
recoiling defect $u_\mu X^\mu = X^0$, and thus the above condition is
imposed at the origin of the co-moving frame target time). In the
linear-dilaton scenario we employed above, this condition translates
to the following metric, which describes the effects of the
interaction of the stringy matter excitation with the D-particle
defect on the surrounding space-time: 
$$g_{\mu\nu}^{\rm matter} =
\eta_{\mu\nu} + \left(e^\Phi -e^{-\Phi}\right)u_\mu u_\nu \ ,\  \Phi =
-u_\mu X^\mu~.$$ The $e^{\Phi}$ correction corresponds to an operator
on the world-sheet of the string with \emph{zero} conformal dimension,
$$ -u_\mu ( - u^\mu - u^\mu ) + 2 = 2 (u_\mu u^\mu + 1) =0~,$$ due to
Eq.~(\ref{fourvelconstr}).  This latter deformation leads to departure
of the associated $\sigma$-model from criticality.

To restore conformal invariance one can follow two approaches, which
we shall only outline here, reserving detailed studies for a future
publication. In view of the landscape scenaria of string theory, both
approaches could lead to acceptable in principle ground states of
strings, which however violate Lorentz invariance, due to the
preferred frame imposed by the recoil velocity field $u_\mu$.

In the first approach, one stays within the Minkowski
$\sigma$-model-frame metric background, and exploits the fact that the
linear dilaton implies a sub-critical string with $Q^2 = u_\mu u^\mu =
-1 < 0$.  This can become conformal (critical) if one uses a {\it
space-like} Liouville mode~\cite{ddk} $\rho$ to ``dress'' the
above-mentioned metric deformations by multiplication with exponential
operators $e^{\alpha_i \rho}$, $i=1,2$, where $\alpha_i$ are the
Liouville ``anomalous'' dimensions.  In the presence of the
world-sheet background charges in the $(\rho, X^\mu)$ extended target
space time~\cite{ddk} of the Liouville-dressed world-sheet theory,
$(|Q|=1, u_\mu )$, which induce world-sheet curvature terms of the
form $$\int_\Sigma d^2\xi (u_\mu X^\mu + \rho)R^{(2)}$$ in the
$\sigma$-model action. The conformal dimension of these operators is
$\alpha_i (\alpha_i + 1)$ for each $i=1,2$.  Restoration of conformal
invariance requires that the total conformal dimension of the
Liouville-dressed deformations is (1,1) in the (holomorphic,
anti-holomorphic) world-sheet sectors. It is straightforward then to
observe that the following dressed operators are conformal, amounting
to the choices $\alpha_i = \pm |Q| = \pm 1$ in the respective
Liouville anomalous dimensions,
\begin{equation}
V_{\rm dress}^{\lambda\nu} u_\lambda u_\nu = \left(e^{-u_\mu X^\mu
-\rho} - e^{u_\mu X^\mu + \rho}\right)\partial X^\lambda {\overline
\partial}X^\nu u_\lambda u_\nu~.
\end{equation} 
These imply a dressed target-space-time metric in the extended
space-time $(\rho, X)$ of the form:
\begin{eqnarray}
&& g_{\mu\nu}^{\rm matter, dressed} (\rho, X) = \nonumber \\ &&
~~~~~~~~~~~~~~~~~~~~~~\eta_{\mu\nu} + \left(e^{\Phi(\rho, X)} -e^{-\Phi(\rho,
X)}\right)u_\mu u_\nu~, \nonumber \\ && g_{\rho \mu} = 0,~~
g_{\rho\rho}=+1, \quad \Phi (\rho, X) =- u_\mu X^\mu - \rho~.
\end{eqnarray}
The extra space-like Liouville mode may thus be given the physical
interpretation of a bulk dimension, in which recoil of the defects
{\it does not take place}.  In this picture, our brane space time is
located at, say, $\rho = 0$.  This is only one example of a consistent
conformal theory.  In general, one may consider non-trivial
$\sigma$-model metrics $G_{\mu\nu}$, in which case the associated
dilatons will have a more complicated space time dependence. In this
respect, one can discuss realistic FRW backgrounds, where however the
dilaton potential (receiving contributions from higher string loops)
plays a crucial r\^ole in the physics~\cite{diamandis,veneziano}.  We
hope to come back to such issues in a forthcoming publication.

In the second approach to restoring conformal invariance, one remains
within the original space time, but departs from the flat Minkowski
background $\sigma$-model metric $\eta_{\mu\nu}$, by allowing a
generic background $G_{\mu\nu}$. The latter can then be determined by
the requirement of the vanishing of the associated Weyl anomaly
coefficients~\cite{tseytlin} ${\tilde \beta_{\mu\nu}^{G,\Phi}}$.
This, in turn, implies~\cite{tseytlin} that the associated
$\sigma$-model renormalisation-group $\beta$-function for the deformed
graviton vertex operator should equal an (infinitesimal) target-space
diffeomorphism variation of the associated background field,
{\it i.e.}, $$\beta^i = -\delta g^i ~~,~~ g^i = (G^{\mu\nu}, \Phi)~.$$ For the
combination of background fields appearing in the expression for the
``metric'' felt by the matter interacting with the D-particle in our
model:
\begin{equation}
g_{\mu\nu}^{\rm matter} \equiv G_{\mu\nu} + \left(e^\Phi
-e^{-\Phi}\right)u_\mu u_\nu~,
\label{combination}
\end{equation} 
one has for the associated world-sheet
$\beta$-functions:
\begin{equation}
\beta_{\mu\nu}^{\rm g^{\rm matter}} = \beta_{\mu\nu}^{\rm G} +
2\beta^\Phi{\rm cosh}(\Phi)u_\mu u_\nu~,
\label{betaf}
\end{equation}
assuming marginal ({\it i.e.}, world-sheet-renormalisation-group-scale
independent) $u_\mu$~\cite{recoil} (see, however, discussion below,
where more general $u_\mu$ may appear).  Requiring conformal
invariance implies that the left-hand-side of Eq.~(\ref{betaf}) must
have the form of the appropriate target space diffeomorphism variation
$\delta g_{\mu\nu}^{\rm matter}$.  From Eq.~(\ref{betaf}) it is
evident that it suffices to consider conformal invariance conditions
for the dilaton $\Phi$ and $\sigma$-model-frame graviton $G_{\mu\nu}$
space time backgrounds, in the presence of the recoil deformations,
which as discussed in Ref.~\cite{recoil}, and will be reviewed below, act
as if they are target space dynamical vector fields coupled to the
metric $G_{\mu\nu}$.  Indeed, in such a case $$\beta_{\mu\nu}^{\rm G}=
-\delta G_{\mu\nu} ~~,~~\beta^\Phi = - \delta \Phi~,$$ and from
Eq.~(\ref{betaf}) it follows trivially that 
\begin{eqnarray}
\beta_{\mu\nu}^{\rm g^{\rm matter}} 
&=& -\delta G_{\mu\nu} - 2{\rm cosh}(\Phi)u_\mu u_\nu \delta \Phi 
\nonumber\\
&=&-\delta \left(G_{\mu\nu} + 2{\rm sinh}(\Phi) u_\mu u_\nu
\right)\nonumber\\
&=& -\delta g_{\mu\nu}^{\rm matter}~,
\end{eqnarray} which guarantees the
world-sheet conformal-invariance of the matter metric
Eq.~(\ref{combination}). 

 In this article we shall not discuss in detail the existence of
solutions to these constraints and their cosmological relevance.  We
only mention that in string models of the type considered in
Ref.~\cite{emn}, where a population of D0-branes exists, it is possible to
consider an isotropic situation, in which the statistical average
$\langle\langle ... \rangle\rangle$ over the D0-branes of the spatial
component of the recoil velocity vanishes, $\langle\langle u _i
\rangle\rangle = 0~, i=1,2,3...$.  Effectively, if one ignores
fluctuations of order $\langle\langle u _i u^i \rangle\rangle$, this
may be considered equivalent to assuming only a temporal non-trivial
component for $u_\mu$, $u_0 \ne 0$, in agreement with the standard
spatially isotropic TeVeS cosmologies~\cite{skordis,liguori}.  The
constraint $u_\mu u^\mu = -1$, then, would imply that $u_0 $ is a
constant, as assumed above, only in cases where the temporal component
of the background metric is $-1$ (as happens, for instance, in
standard FLRW cosmologies in the appropriate
time frame).  In general, however, one may consider other frames, or
more general situations, for instance the conformal time ones
of Refs.~\cite{skordis,liguori}, in which the temporal component of the
$\sigma$-model metric will not be $-1$.  In such cases, $u_\mu$ are
not constants, but depend on the dilaton, the scale factor {\it etc.},
and the conformal invariance conditions get more complicated. We
postpone a detailed discussion of such a situation for a forthcoming
publication.

In either of the above ways of restoring the conformal invariance of the 
$\sigma$-model, 
we note the existence of two metrics, which is reminiscent of TeVeS theory,
Eq.~(\ref{mattermetric}). One, is the $\sigma$-model background metric,
metric, $G_{\mu\nu}$, which is related to an Einstein-frame metric
$g^E_{\mu\nu}$ via~\cite{aben}: $g^E_{\mu\nu} = e^{-4\Phi/(d-2)}G_{\mu\nu}$
in d-dimensional space-time.
The other, is the metric $g_{\mu\nu}^{\rm matter}$ 
describing the distortion of the space-time
surrounding the D-particle defect, as a result of its interaction
with stringy matter: \be g_{\mu\nu}^{\rm matter} = e^{4\Phi/(d-2)}
g^E_{\mu\nu} + \left(e^\Phi -e^{-\Phi}\right)u_\mu u_\nu~.
\label{mattermetric3}
\ee The reader is invited to compare Eqs.~(\ref{mattermetric}) and
(\ref{mattermetric3}); there are differences in the coefficients of
the scalar field in the various exponentials, but the basic
qualitative bi-metric features, are common.  The dynamical scalar
field of the TeVeS theory is thus played by the dilaton field in this
model. The latter can also be responsible for yielding
quintessence-like dark energy
contributions~\cite{diamandis,veneziano,emmncosmo}.  We observe that
for $d=6$, one encounters a precise analogy with TeVeS models, which
implies a $\Phi$-independent electromagnetic fine-structure
constant~\cite{teves}.  The velocity field $u_\mu$, which is subject
to the constraint Eq.~(\ref{fourvelconstr}) is \emph{not} directly related
to the dynamical vector field $A_\mu$ of the TeVeS theory, subject to
the constraint, Eq.~(\ref{constrvec}).

There is a dynamical gauge background field in the D-particle recoil
model, which upon $T$-duality (which exchanges Neumann and Dirichlet
boundary conditions) is related to the recoil
deformation~\cite{recoil} : \be {\cal A}_\mu = u_\mu \Phi \Theta
(\Phi) ~, \qquad \Phi = -u_\mu X^\mu~,
\label{gauge}
\ee
with Lagrangian of the Born-Infeld type:  \be \frac{1}{g_s}\sqrt{{\rm
det}\left(\eta_{\mu\nu} + 2\pi \alpha ' g_s^2 F_{\mu\nu} 
\right)} \ni  \alpha' g_{\rm s}^3 F_{\mu\nu}F^{\mu\nu};
  \label{bi}
  \ee $F_{\mu\nu}({\cal A}) $ is the (Abelian) field strength. This
holds for flat backgrounds.  In general, for non-trivial 
$\sigma$-model-frame metric backgrounds $\eta_{\mu\nu} \to 
G_{\mu\nu}$ in (\ref{bi}), 
and there are also overall dilaton
exponential factors, leading to relaxation dark energy 
terms~\cite{diamandis,emn,veneziano}.

In a galactic region, where we expect a statistically significant
population of D-particles, there is a distribution of recoil
velocities, thus one has to average Eq.~(\ref{gauge}) over such
three-velocity distributions.  We make the physically plausible
assumption that such distributions are \emph{isotropic} in
space~\cite{emn}, $\langle\langle v_i \rangle\rangle = 0$.  In this
sense, the so-averaged gauge field depends solely on the target time
$X^0>0$, and is of the form of the cosmological TeVeS field.
In view of Eq.~(\ref{fourvelconstr}), the gauge field obeys a
\emph{gauge fixing} condition 
${\cal A}_\mu {\cal A}^\mu = -\Phi^2$,
which plays the r\^ole of the constraint Eq.~(\ref{constrvec}).
The indices are raised and lowered by the
Einstein-frame metric $g_{\mu\nu}^E$. In the cosmological context of
the TeVeS model, one may assume for $g_{00}$ the form $g_{00}^E =
a(\tau)^2 e^{-2\Phi}\eta_{00}$, with $a$ the scale factor of the
universe.  This implies that $u^\mu = {\rm d}X^\mu/{\rm d}\tau$, 
where $X^\mu$ are
$\sigma$-model-frame coordinates, and $\tau$ is a D-particle co-moving
frame (proper) time. The co-moving cosmic time is then defined in the
Einstein frame of the string by~\cite{aben}: ${\rm d}\tau = a e^{-\Phi}
{\rm d}X^0 $, {\it i.e.},  $u^0 = a^{-1} e^\Phi $. Thus, like the TeVeS vector
field, the Born-Infeld gauge field,
describing the interaction of stringy matter with D-particle
space-time defects, has the form 
\begin{equation}
\langle \langle {\cal A}_0 \rangle\rangle = a e^{-\Phi} \Phi \Theta (\Phi)~.
\end{equation}
From  Eq.~(\ref{bi}) we observe
that the r\^ole of the parameter $K$ of the TeVeS theory (the
coefficient of the Maxwell-like kinetic term for the vector field) is
played here by the $ g_{\rm s}^3 $, which is $<
1$, for weak string coupling.  For low-values of
$K$, there is enhanced growth in the density
perturbations~\cite{liguori}.

We next remark on the special r\^ole of neutrinos in inducing
non-perturbative cosmological-constant-type contributions to the
vacuum energy in this picture. Neutrinos, as being electrically
neutral, could be of the type of string excitations that interact with
D-particles, leading to the existence of Lorentz-violating vector
fields, Eq.~(\ref{gauge}), and the associated cosmological
instabilities.
The important point to notice~\cite{mavsar} is that ``flavor'' is not
\emph{necessarily conserved} in such interactions.  After the capture
by the D-particle defect, the emerging stringy matter excitation could
have a different flavor than what it had initially.  Thus, the
D-particle populations in galaxies act as a ``medium'' inducing flavor
oscillations, in analogy with the celebrated
Mikheev-Smirnov-Wolfenstein (MSW) effect~\cite{msw}.  Following
Refs.~\cite{mavsar,barenboim} we represent the ground state of the
space-time with D-particle defects with which neutrinos interact as a
Fock-space ``flavor'' vacuum $|0\rangle _f$, introduced in
Ref.~\cite{vitiello} in order to discuss quantisation of
field-theories with mixing.  Since in galaxies there is abundance of
electrons (stable), there is a significant contribution to the vacuum
energy coming from oscillations $\nu_e \leftrightarrow \nu_\mu$, since
it is the electron current that couples to the muon neutrino current
in Standard Model interactions.  Thus, the dominant contributions to
the vacuum energy can be computed in a two-generation neutrino
model~\cite{vitiello,barenboim}.  Considering a cosmological
space-time, say of FLRW type, and computing the average of the
neutrino stress tensor with respect to the flavor vacuum $_f\langle 0
|T_{\mu\nu}| 0 \rangle_f$, one finds that, in contrast to the usual
mass-eigenstate Lorentz-invariant-vacuum case, where this quantity
vanishes, there is a non-trivial, non-perturbative contribution to the
vacuum energy~\cite{vitiello}: 
\begin{equation}
\rho_{\rm vac} = \frac{2}{\pi}{\rm
sin}^2\theta \int_{0}^{K_0} d^3k (\omega_{k,1} + \omega_{k,2})|V_{\vec
k}|^2~,
\end{equation} $K_0$ a momentum cutoff, determining the
relevant low-energy degrees of freedom, and
\begin{equation}
|V_{\vec k}|^2 \sim (m_1-m_2)^2/(4|{\vec k}|^2)~,
\end{equation} for large momenta,
the ``flavor'' condensate.
A consistent, and physically relevant choice of the cutoff 
is~\cite{barenboim} 
\begin{equation}
K_0 \sim m_1 + m_2~.
\end{equation} 
This choice is compatible
with our D-particle model since it implies that only the infrared
neutrino modes, with momenta less than the typical mass scales $m_1 +
m_2$, feel mostly the ``D-particle medium'' effects (being slow, they
have more time to interact with them).  Thus,
~\cite{barenboim,mavsar}
$$\rho_{\rm vac}
\sim
\frac{2}{\pi}{\rm sin}^2\theta (m_1 - m_2)^2 (m_1 + m_2)^2 \sim
{2\over\pi}{\rm sin^2}\theta (\Delta m^2_{12})^2~.
$$ For $\Delta m^2_{12} \sim 7 \times 10^{-5} $~eV$^2$, and ${\rm
sin}^2\theta \sim .3$, which are the measured values from atmospheric
neutrino experiments, this leads to a dark-energy contribution from
neutrinos
\be 
\Omega_{\rm \Lambda}^{\nu_{\rm mixing}} \sim 0.24~. 
\label{omegal}
\ee The reader should keep in mind that there are additional
time-dependent dark-energy contributions coming from the dilaton
quintessence field~\cite{diamandis,veneziano}, which are not fully
understood, as they include string loop corrections.

It is claimed~\cite{skordis} that the correct position of the CMB
peaks is obtained if neutrino masses are of order 2 eV, their ``dark
matter'' contribution is $\Omega_\nu \sim 0.15- 0.17$, and the total
$\Omega_\Lambda\sim 0.80-0.78$, respectively, with 5\% ordinary
matter. These features can be accommodated, if necessary, in our
scenario.  The presence of massive neutrino dark matter in galactic
centres, would also contribute to the modification of the rotational
curves of galaxies, on equal footing with the vector field.  A
detailed phenomenology of our model is left for future study.

A last but important point  is that the
D-particle medium consists of \emph{supersymmetric} D0-brane defects,
which are such that there is no contribution to the vacuum energy if
interactions (recoil ``movements'') with stringy matter are
ignored. The D-particles are a sort of BPS states, which
experience a zero net-force between them. Thus, the vacuum stress
tensor of these supersymmetric defects vanishes and  one avoids
large isocurvature perturbations, which was a problem of other
non supersymmetric defects.
Isocurvature perturbations are induced by the presence of ``seeds''
(any non-uniformly distributed form of energy, which contributes only
a small fraction to the total energy density of the universe and which
interacts with the cosmic fluid only gravitationally). Such
perturbations are generated continuously and evolve according to
inhomogeneous linear perturbation equations. The randomness of the
non-linear seed evolution, which sources the perturbations, can
destroy the coherence fluctuations in the cosmic
fluid~\cite{ms1}. Moreover, if the seeds have non-linear dynamics,
then the distribution of  anisotropies leads to
non-Gaussian statistics. Incidentally, small (unobservable) deviations
from Gaussian statistics can also appear if the initial state of the
field, responsible for the origin of fluctuations, is in a non-vacuum
initial state~\cite{jmms}.  Isocurvature perturbations and strong
deviations from Gaussian statistics plague topological
defect~\cite{td}, as well as seed models in the context of
pre-big-bang cosmology~\cite{ven}. Severe constraints have been
imposed~\cite{constr} to the contribution of isocurvature
perturbations to the CMB temperature anisotropies. In our model,
D-particles could also lead to non-Gaussian signatures, due to
$\langle\langle u_i^2\rangle\rangle\ne 0$, which  remain
undetectable.  In our case, however, the presence of non-trivial dilaton fields 
may imply in general strong isocurvature perturbations, whose suppression
amounts to severe constraints on model building. We hope to come back to 
such issues
in the future.

In this work we have attempted to present models from string theory,
leading to  Lorentz-violating isotropic vector
fields, and thus to a TeVeS theory.  We have argued that one such
model involves supersymmetric D-particles as a gravitational ground
state ``medium'' on which stringy matter propagates. 
The interactions of matter with the defect involve ``capture'' of the 
matter string by the D0-brane, which leads 
to a different metric felt by the matter
string as compared with the space-time background metric. The difference
involves deformations by the recoil velocity field. 
We have presented arguments supporting the consistency of this bi-metric 
theory with world-sheet conformal invariance, which therefore makes it an
acceptable string ground state. We have also argued that 
neutrinos are among the matter species that
can interact in the above topologically non-trivial way (``capture'') 
with the D0 brane
defects, 
and we have discussed the appearance of 
novel, non-perturbative contributions to the dark energy
of the Universe 
as a result of the neutrino-D-particle interactions 
in such a model.  

A final remark we would like to make 
is that the interactions of stringy matter with D-particles break,
of course, the target-space supersymmetry of the vacuum, but this
breaking is rather a \emph{supersymmetry} obstruction~\cite{emn}, in
the sense of yielding a non-supersymmetric spectrum of excitations.
The breaking will be of order of the recoil velocity fluctuations
$\langle\langle u_i^2 \rangle\rangle $, and thus for realistic
situations very small.  Hence, any phenomenologically realistic
supersymmetry breaking on the D-brane world should be obtained from
other means.

\vskip.05truecm
\acknowledgments This work is partially supported by the European
Union through the Marie Curie Research and Training Network
\emph{UniverseNet} (MRTN-CT-2006-035863).


\end{document}